\documentstyle[twocolumn,aps,epsf]{revtex}

\title{Hopping motion of lattice gases through nonsymmetric potentials
  under strong bias conditions}

\author{K.W. Kehr$^1$ and Z. Koza$^{1,2}$}
\address{$^1$Institut f\"ur Festk\"orperforschung, Forschungszentrum
  J\"ulich GmbH,  D-52425 J\"ulich, Germany,\\
$^2$Institute of Theoretical Physics, University of
Wroc{\l}aw, pl. Maxa Borna 9, PL-50204 Wroc{\l}aw,
Poland}

\date{\today}

\begin{document}
\maketitle

\begin{abstract}
The hopping motion of lattice gases through potentials without mirror-reflection
symmetry is investigated under various bias conditions. The model of 2 particles on a ring 
with 4 sites is solved explicitly; the resulting current in a sawtooth potential is discussed.
The current of lattice gases in extended systems consisting of periodic repetitions of
segments with sawtooth potentials is studied for different concentrations and values of
the bias. Rectification effects are observed, similar to the single-particle case.  
A mean-field approximation for the current in the case of strong bias acting against the
highest barriers in the system is made and compared with numerical simulations. 
The particle-vacancy symmetry of the model is discussed.
\end{abstract}

\section{Introduction}
The motion of particles in potentials that do not have mirror reflection  
symmetry has attracted much attention
in the last years for several reasons. The interest extends from fundamental problems 
concerning the validity of the Second Law of Thermodynamics \cite{feyn,magn1}
to applications in biological \cite{HMS,stark,MC,rda,JAP}
and chemical systems \cite{ZC}, as well as for solid-state devices \cite{Letal,Setal}. 
Major efforts have been devoted to an understanding of molecular motors, where proteins
move in nonsymmetric potentials under the influence of stochastic and/or other forces.
One specific observation for transport in nonsymmetric potentials is the
possibility of rectification effects if the forces on the particles are beyond the
regime where linear-response theory is applicable \cite{KMWD}. Rectification effects have been 
discussed in continuous \cite{CM} as well as in hopping systems \cite{stark,KMWD}. 
If applications of effects
of particle motion in nonsymmetric potentials are envisaged then the question 
arises as to the influence of many-particle effects. The limit of
single-particle motion is rarely realized; in real systems many 
particles are present that compete about the sites that can be occupied.
Many-particle effects have been studied in continuous nonsymmetric 
periocic potentials in Ref.\cite{DV}, where interesting dependencies of the current on 
particle concentration and size were found. In this paper 
we will investigate hopping motion of lattice-gas
particles in 
nonsymmetric hopping potentials under the influence of strong bias. We utilize the
simple site exclusion model where
multiple occupancy of sites is excluded and direct our attention to nonlinear effects 
on the particle current.
 
The stationary current of a single particle performing a hopping motion in a nonsymmetric
potential under an arbitrary bias is known exactly \cite{KMWD}. The 
calculation of the stationary current of 
site-exclusion lattice gases in nonsymmetric potentials that lead to 
rectification effects in the single-particle case is a difficult problem. Extensive
work has been devoted to the asymmetric site exclusion process including the
totally asymmetric site exclusion process (TASEP) where
the particles can only hop in one direction, corresponding to very strong bias. The
case of uniform hopping potentials is now well understood \cite{schuetz}, but the case of nonuniform
potentials is not generally solved. Recent work has been devoted to the TASEP with disordered
potentials \cite{JL,schuetz2,TB1,TB2,GS,KP}.
For the general asymmetric case one has to resort to numerical
simulations; we are going to present simulation results for the stationary current
of lattice-gas particles in nonsymmetric potentials, for various concentrations and values 
of the bias. 

Nonetheless, some analytical treatment can be given.
First, the case of very small periodic systems can be treated explicitly:
the motion of 2 particles
on a ring of period 4 can be solved by elementary means. Although this is a very simple 
system, conclusions can be drawn in the limit of very strong bias that are of interest
for the totally asymmetric site exclusion process. 
The nonlinear current of site-exclusion lattice gases in extended systems with periodic
repetitions of nonsymmetric segments can be
derived in a mean-field approximation for strong bias conditions. 
Interesting symmetry properties have been pointed out 
for the TASEP in disordered hopping potentials \cite{TB2,GS,KP}. While a particle-vacancy 
symmetry is also present in our model, the case of inversion of the bias direction is
different here.

In the following section the hopping motion of 2 particles on
a ring of period 4 is solved and analyzed. In Sec.\ III a mean-field approximation for 
the stationary current
of lattice gases under strong bias in nonsymmetric hopping potentials
is presented and compared with simulation results in a sawtooth potential. 
The symmetry properties of the model are discussed in Sec.\ IV and
concluding remarks are given in Sec.\ V.

\section{Two Particles on Ring with Four Sites}

\subsection{Solution
             of the stationary master equations}

\begin{figure}
  \epsfxsize3.5in 
  \epsfbox{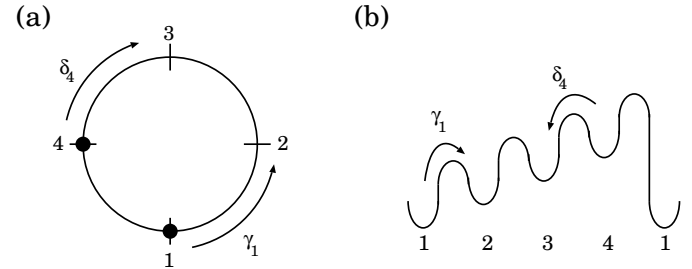}
  \caption{
    \label{Fig1}   
(a)Ring of 4 sites with 2 particles. (b) Sawtooth potential with period 4 
without bias ($b=1$) with 2 transition rates indicated.
  }
\end{figure}

A very simple yet nontrivial model is given by a ring with 4 sites and 2 particles, 
cf. Fig.\ \ref{Fig1}. The basic quantities for the description of the system are the joint 
probabilities
$P(i,j;t) \:(i\ne j)$ of finding one particle at site $i$ and the other particle at site $j$,
at time $t$, for specified initial conditions. Since the particles are considered as
indistinguishable, $P(i,j;t) = P(j,i;t)$. There are 6 different joint two-particle probabilities 
on the ring with 4 sites (generally $L(L-1)/2$ on rings with $L$ sites). Higher-order joint
probabilities do not occur for 2 particles. 

The probabilities $P(i;t)$ of finding a particle at site $i$ at time $t$ are given by
\begin{equation}
  \label{pp}
  P(i;t) = \sum_{j\ne i} P(i,j;t)
\end{equation}
For two particles they are normalized to 
\begin{equation}
  \label{np1}
  \sum_{i=1}^4 P(i;t) =2 \:.
\end{equation}
This condition implies 
\begin{equation}
  \label{np2}
  \sum_{i < j} P(i,j;t) = 1 \:.
\end{equation}
 The master equations for the joint probabilities are easily written
down,
{ 
 \begin{eqnarray}
   \label{ME}
   \frac{d}{dt}P(1,2;t) &=& \delta_3P(1,3;t)+\gamma_4P(2,4;t)
-(\gamma_2+\delta_1)P(1,2;t) \nonumber \\
   \frac{d}{dt}P(2,3;t) &=& \delta_4P(2,4;t)+\gamma_1P(1,3;t)
-(\gamma_3+\delta_2)P(2,3;t) \nonumber \\
   \frac{d}{dt}P(3,4;t) &=& \delta_1P(1,3;t)+\gamma_2P(2,4;t)
-(\gamma_4+\delta_3)P(3,4;t) \nonumber \\
   \frac{d}{dt}P(1,4;t) &=& \delta_2P(2,4;t)+\gamma_3P(1,3;t)
-(\gamma_1+\delta_4)P(1,4;t) \nonumber \\
\frac{d}{dt}P(1,3;t) &=& \gamma_2P(1,2;t)+\delta_2P(2,3;t)+\gamma_4P(3,4;t)
  \nonumber \\ 
  &+& \delta_4P(1,4;t) - (\gamma_1+\delta_1+\gamma_3+\delta_3)P(1,3;t) 
  \nonumber \\
\frac{d}{dt}P(2,4;t) &=& \delta_1P(1,2;t)+\gamma_3P(2,3;t)+\delta_3P(3,4;t)
  \nonumber \\
  &+& \gamma_1P(1,4;t) - (\gamma_2+\delta_2+\gamma_4+\delta_4)P(2,4;t).
 \nonumber \\
\end{eqnarray}
}
The sum of the 6 master equations leads to the conservation law
\begin{equation}
  \label{cons}
  \frac{d}{dt}[\sum_{i<j} P(i,j;t)] = 0 \:,
\end{equation}
consistent with the relation (\ref{np2}) given above.

We are interested in the stationary solution of the system of master equations (\ref{ME}). 
The stationary values $P(i,j;t \rightarrow \infty)$
will be denoted by $P_{ij}$. The 
stationary joint probabilities for adjacent sites, e.g. $P_{12}$, can all be expressed 
by the stationary joint probabilities $P_{13}$ and $P_{24}$. For instance, the first
line of Eq.(\ref{ME}) yields
\begin{equation}
  \label{elim}
  P_{12} = \frac{1}{\gamma_2+\delta_1}(\delta_3 P_{13}+\gamma_4 P_{24}) \:.
\end{equation}
Three analogous relations follow from (\ref{ME}); they can be obtained by cyclically
increasing the indices in Eq.(\ref{elim}). If the joint probabilities for adjacent sites
are eliminated from the stationary master equations, two homogeneous equations remain
which are equivalent. We write this equation as 
\begin{equation}
  \label{lin1}
  a_{11}P_{13}+a_{12}P_{24}=0
\end{equation}
with the coefficients
\begin{eqnarray}
  \label{co1}
a_{11} &=& -(\frac{\delta_1\delta_3}{\gamma_2+\delta_1}+\frac{\gamma_1\gamma_3}{\gamma_3+\delta_2} 
+\frac{\delta_1\delta_3}{\gamma_4+\delta_3}+\frac{\gamma_1\gamma_3}{\gamma_1+\delta_4}) \nonumber \\
a_{12} &=& \frac{\gamma_2\gamma_4}{\gamma_2+\delta_1}+\frac{\delta_2\delta_4}{\gamma_3+\delta_2} 
+\frac{\gamma_2\gamma_4}{\gamma_4+\delta_3}+\frac{\delta_2\delta_4}{\gamma_1+\delta_4} \:.
\end{eqnarray}

The second equation for $P_{13}$ and $P_{24}$ is obtained from the normalization condition
Eq.(\ref{np2}), after elimination of the joint probabilities of adjacent sites. It reads
\begin{equation}
  \label{lin2}
  a_{21}P_{13}+a_{22}P_{24} = 1 \:.
\end{equation}
with the coefficients
\begin{eqnarray}
  \label{co2}
a_{21} &=& \frac{\delta_3}{\gamma_2+\delta_1}+\frac{\gamma_1}{\gamma_3+\delta_2} 
+\frac{\delta_1}{\gamma_4+\delta_3}+\frac{\gamma_3}{\gamma_1+\delta_4}+1 \nonumber \\
a_{22} &=& \frac{\gamma_4}{\gamma_2+\delta_1}+\frac{\delta_4}{\gamma_3+\delta_2} 
+\frac{\gamma_2}{\gamma_4+\delta_3}+\frac{\delta_2}{\gamma_1+\delta_4}+1 \:.  
\end{eqnarray}
The solution of the two linear equations is
\begin{eqnarray}
  \label{sol}
  P_{13} &=& \frac{-a_{12}}{a_{11}a_{22}-a_{12}a_{21}}\:, \nonumber \\
  P_{24} &=& \frac{a_{11}}{a_{11}a_{22}-a_{12}a_{21}} \:.
\end{eqnarray}
Since the joint probabilities for adjacent sites are obtained from the $P_{13}$, $P_{24}$,
and the one-site stationary probabilities $P_i \equiv P(i;t\rightarrow \infty)$ from Eq.(\ref{pp}),
Eq.(\ref{sol}) represents the complete solution of the stationary problem.

We derive the stationary current in the system by considering the bond connecting 
sites 1 and 2. The stationary current is given by 
\begin{equation}
  \label{curr1}
  J=\gamma_1(P_1-P_{12})-\delta_2(P_2-P_{12})
\end{equation}
The joint probabilities in Eq.(\ref{curr1}) ensure exclusion of double occupancy of sites.
Using Eq.(\ref{pp}) the current is expressed in terms of the joint probabilities,
\begin{equation}
  \label{curr2}
  J=\gamma_1(P_{13}+P_{14})-\delta_2(P_{23}+P_{24}) \:.
\end{equation}
Insertion of the stationary solution for the joint probabilities gives 
\begin{equation}
  \label{curr3}
  J = \frac{\gamma_1+\gamma_3+\delta_2+\delta_4}{(\gamma_1+\delta_4)(\gamma_3+\delta_2)}
  (\gamma_1\gamma_3P_{13}-\delta_2\delta_4P_{24}) \:.
\end{equation}
The current may also be derived by considering the other bonds of the ring. Two equivalent forms
of the current result; the second (equivalent) form reads 
\begin{equation}
  \label{curr4}
  J = \frac{\gamma_2+\gamma_4+\delta_1+\delta_3}{(\gamma_2+\delta_1)(\gamma_4+\delta_3)}
  (\gamma_2\gamma_4P_{24}-\delta_1\delta_3P_{13}) \:.
\end{equation}
It can be shown that the current vanishes if the right and left transition rates fulfill
the following condition,
\begin{equation}
  \label{det}
  \gamma_1\gamma_2\gamma_3\gamma_4 = \delta_1\delta_2\delta_3\delta_4
\end{equation}
corresponding to a detailed balance relation over the ring.

\subsection{Solution for the sawtooth potential}

The sawtooth potential including bias on a 4-site ring is defined by choosing 
\begin{eqnarray}
  \label{DefSawtooth}
    \gamma_1 &=& \gamma_2 = \gamma_3 = \gamma_4 = b\gamma,
       \nonumber\\
    \delta_1 &=& b^{-1}\gamma^4 ,
       \nonumber \\
    \delta_2 &=&  \delta_3 = \delta_4 = b^{-1}, 
\end{eqnarray}
where $b$ represents the  bias and $\gamma < 1$ is a constant
representing a transition rate to the right in the absence of a bias, cf. Fig.1(b).
Note that the right transition rates are explicitly multiplied by
the bias factor $b$ and the left transition rates by $b^{-1}$,
respectively. Physically, $b = \exp (\Delta U/2k_BT)$ where $\Delta U$ represents the
potential drop between two neighboring sites under the influence of the bias. 
For $b=1$ the system satisfies the detailed balance
condition and the current $J$ vanishes. In what follows the current
obtained in a system with $M$ particles will be denoted as $J_M$.

\begin{figure}
  \epsfxsize3.5in 
  \epsfbox{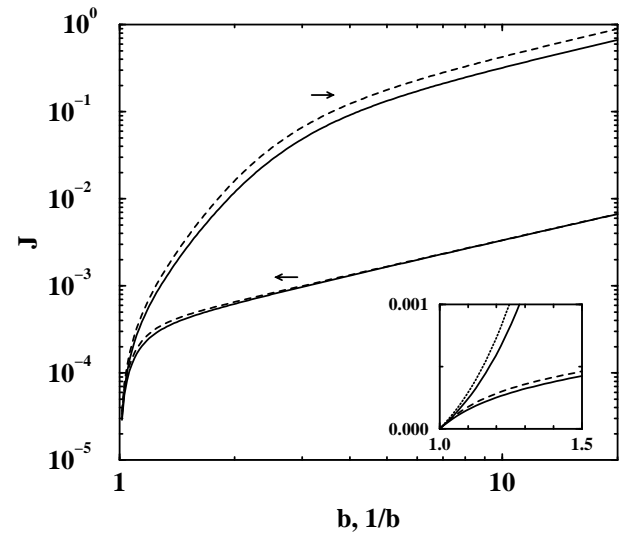}
  \caption{
    \label{FigK2}   
Particle current $J$ (arbitrary units) as a
function of the dimensionless bias parameter. Upper curves: bias to the right, ordinate 
indicates $b$; lower curves: bias to the left, ordinate indicates $1/b$. Dashed lines:
2 particles on the ring with 4 sites; Full lines: single particle on the ring.
Inset: behavior for small bias (linear axes).
  }
\end{figure}

In Fig.\ \ref{FigK2}  we present a plot of $J_1$ and $J_2$ as functions of the
bias $b$ for the ring with 4 sites  and $\gamma = \exp(-2)$.  The result for the
two-particle system was obtained using Eq. (\ref{curr4}), and for a
single-particle system we employed the exact solution derived in
Ref. \cite{KMWD}.  We can see that the behavior of the currents of
one- and two-particle systems are qualitatively similar. Of course,
the current $J_{2}$ of two particles is larger than the one-particle
current $J_{1}$. The inset shows the behavior of the current for smaller
bias. The curves for the bias to the right and to the left become equal
in the limit $b \rightarrow 1$, i.e., in the linear-response regime,
for two particles on the ring with 4 sites, and also for one particle
on this ring. However, the two-particle current is about $17\: \% $
larger than the one-particle current. 

In the case of a strong bias to the right, $b\gg1$, the two-particle
current $J_2$ differs from $J_1$ by a constant factor.  For the
sawtooth potential this behavior can be understood as follows. If $b
\gg 1$, only transitions to the right are important, and backward
transitions can be neglected. In our model the transition
rates to the right are all equal, $\gamma_i = b\gamma$ for
$i=1,\ldots,4$. Hence for $b \gg 1$ all stationary site occupation
probabilities become equal, $P_i = 1/2$ for the 4-site ring and
$P_i=2/L$ for a ring with L sites. In the limit of a strong bias to
the right all stationary joint probabilities also become equal, i.e.,
$\forall_{i,j}$ $P_{ij} = 1/6$ for the 4-site ring and, generally,
$P_{ij} = 2/L(L-1)$ (see Ref. \cite{Derrida}).  Using expression
(\ref{curr1}) we thus expect that for $b\gg 1$
\begin{equation}
  \label{curr5}
  J \approx b\gamma \left[\frac{2}{L}-\frac{2}{L(L-1)} \right].
\end{equation}
For $L=4$ there is thus $J_{2}=b\gamma /3$, which should be compared
with the single-particle current $J_{1}=b\gamma /4$. Similarly, in the
general case of an $L$-site ring we have
\begin{equation}
  \label{RightLimit}
  \lim_{b\to\infty} \frac{J_{2}}{J_{1}} = \frac{2(L-2)}{L-1}.
\end{equation}
For the 4-site ring this limiting behavior can be easily derived from
the exact formula (\ref{curr4}). Actually, for $L=4$, $\gamma =
\exp(-2)$, and $b = 10$ there is $J_{2}/J_{1} \approx 1.3337$, in agreement
with the above considerations.

Figure\ \ref{FigK2}  also shows that $J_{2}$ becomes almost identical to $J_{1}$
in the case of a strong bias to the left, $b \ll 1$. To understand
this phenomenon assume that $\gamma \ll 1$, so that $\delta_1 \ll
\delta_2 = \delta_3 = \delta_4$, i.e., site 1 acts as a ``bottle-neck''.  If
$b \ll 1$ the particles are driven against the high barrier at site 1,
which has a relatively very small transition rate $\delta_1$ to the left.  The
second particle on site 2 has to wait until the first particle has
jumped over the high barrier, and only then can it make
an attempt to jump over that barrier. Soon after the first particle
has managed to pass the bottleneck at site 1, the second particle will jump
from site 2 to 1 and  the first
particle will quickly line up behind the second particle, waiting for
it to jump over the high barrier.
Consequently,
the current becomes practically equal to that of a single-particle system.
It is evident that in the limit of a large bias to the left the system
behaves as a TASEP on a ring with one defect. If the defect is
characterized, in a discrete-time dynamics, by the transition
probability $p \ll 1$, the current of $M$ particles on a ring with
$L$ sites ($M < L$) will approach the one-particle current. The
above reasoning is confirmed by an explicit calculation of the current
$J_{2}$ in the limit $b\to 0$. Using (\ref{curr4}) we conclude, after
some algebra, that $J_{2}
\approx 2b^{-1}(1+\gamma^4)(5\gamma^4 + 2\gamma^{-4} + 5)^{-1}$.
Since for a single-particle system the current $J_{1}$, for $b \ll 1$,
is approximately equal $b^{-1}\gamma^4(1+3\gamma^4)$ (see Ref.\
\cite{KMWD}), we find that
\begin{equation}
   \label{LeftLimit}
   \lim_{b\to 0} \frac{J_{2}}{J_{1}} = 
       \frac{2(1+\gamma^4)(1+3\gamma^4)}{5\gamma^8 + 5\gamma^4 + 2}.
\end{equation}
For  $\gamma\to 0$, i.e., for a growing asymmetry of the sawtooth
potential,  this limit actually approaches 1. In
particular, for the value of $\gamma = \exp(-2)$ used in Fig.\ 2 there
is $\lim_{b\to 0} {J_{2}}/{J_{1}} \approx 1.0005$.

Note, however, that in contrast to the case $b\gg1$, for $b\ll 1$
the current depends on the parameter $\gamma$ characterizing the
inhomogeneity of the sawtooth potential. In particular, for $\gamma =
1$, which corresponds to a fully homogeneous system, $\lim_{b\to 0}
{J_{2}}/{J_{1}} = 4/3$. Actually, for $\gamma = 1$, the ratio
${J_{2}}/{J_{1}}$ equals $4/3$ irrespective of the bias $b$ (see
\cite{Derrida}).

\section{Extended Nonsymmetric Potentials}

\subsection{The model}

In this section lattice gases in extended potentials are considered that consist of 
periodic repetitions of nonsymmetric segments. First the situation of very strong bias is discussed
and a mean-field approximation 
is given for the case where the particles experience periodically
arranged high barriers. The analytical results are then compared with 
numerical simulations of the motion of lattice-gas particles
in nonsymmetric hopping potentials for different concentrations and under various bias conditions.
The hopping potential that is used in in this section is the sawtooth potential as
shown in Fig.\ 1(b), except that it is periodically repeated with period
$L$. The nearest-neighbor transition rates from site $l$ to $l\pm 1$ are 
$\Gamma_{l,l\pm 1}$. As a short notation 
we use $\gamma_l \equiv \Gamma_{l,l+1}$ for the ``right'' and $\delta_l\equiv \Gamma_{l,l-1}$
for the ``left'' transition rates. Without additional bias, the transition rates between
neighbor sites fulfill detailed balance. 
Bias is introduced by multiplying all right transition rates by $b$, $\gamma_l \rightarrow
b\gamma_l$, and all left transition rates by $b^{-1}$,  $\delta_l \rightarrow
b^{-1}\delta_l$.  

The linear chain on which the model is defined shall have $N=\nu L$ sites where we 
consider $\nu \gg 1$ in this section. Periodic boundary conditions are introduced
and the sites are occupied by $M$ particles. The concentration is then $\rho =M/N$.
Multiple occupancy of the sites is excluded; no further interactions of the particles 
are taken into account.

\subsection{Strong bias}

\subsubsection{The case $b \gg 1$}
For $b\gg 1$ we can apply essentially the same reasoning as in the
case of the 2-particle system considered in Sec.\ II. In this limit 
 transitions to the left are so rare that they can be ignored and the
system essentially behaves like a TASEP with transitions $\gamma_i =
b\gamma$, $i = 1,\ldots,N$. The current for such a system reads
\cite{Derrida}
\begin{equation}
  \label{J_TASEP}
  J =b\gamma  M\frac{N-M}{N-1}\:.
\end{equation}
For large system sizes $N \gg 1$ this formula can be rewritten as
\begin{equation}
  \label{J_TASEP_RO}
  J(\rho) =b\gamma  \rho(1-\rho)\:.
\end{equation}

\subsubsection{The case $b \ll 1$}
For $b\ll 1$ we can neglect transition rates to the right, and so the
system behaves like a TASEP with transition rates $\delta_i =
b^{-1}\gamma^4$ if $i = 1 \:(\mbox{mod }L)$ and $\delta_i = b^{-1}$
otherwise. If additionally $\gamma = 1$, all $\delta_i$ are equal to
each other and the current is given simply by (\ref{J_TASEP}).

A more complicated situation appears for $\gamma \ll 1$, a condition
which will be assumed henceforth. In this case sites
$i=1,L+1,\ldots,N-L+1$ act on the flow of particles as ``bottlenecks'',
for the mean time necessary to leave them is much larger than the time
to leave any other site. Therefore the system, which consists of $\nu$
similar segments of length $L$, effectively behaves like a ring made up
of $\nu$ similar ``boxes'', each able to contain up to $L$ particles. 
A transition from a segment $j$ to $j-1$ occurs with a rate
$b^{-1}\gamma^4$ irrespective of the number of particles in each of
the segments, provided, of course, that there is at least one particle in
segment $j$ and at most $L-1$ particles in segment $j-1$. 

Let $Q_n$ denote the probability that in the steady state there are
$n$ particles in a given segment ($n=0,\ldots,L$).  Let $Q_{m,n}$
denote the joint probability of finding, in the steady state, $0 \le m
\le L$ particles at a given segment $j$ and $0 \le n \le L$ particles
at $j+1$. Of course $Q_n$ and $Q_{m,n}$ do not depend on $j$, and the $Q_n$
satisfy 
\begin{eqnarray}
  \label{Prob=1}
   \sum_{n=0}^{L} Q_n &=& 1,  \\
  \label{Rho}
   \sum_{n=0}^{L} nQ_n &=& L\rho.
\end{eqnarray}

Let us assume a mean-field approximation: $Q_{m,n} = Q_m Q_n$. In the
stationary state the mean number ${\cal N}_n$ of segments occupied by
$n$ particles does not depend on time.  As the particles hop between
segments, ${\cal N}_n$ can decrease when one of the particles jumps
from or to a segment occupied by $n$ particles. The corresponding
rates are $Q_n(1-Q_L)$ and $Q_n(1-Q_0)$, respectively. The number of
segments containing $n$ particles can also increase owing to jumps
ending at segments containing $n-1$ particles or originating at
segments with $n+1$ particles; the corresponding transition rates are
$Q_{n-1}(1-Q_0)$ and $Q_{n+1}(1-Q_L)$, respectively.  Consequently,
the appropriate balance conditions read
\begin{eqnarray}
  \label{Balance1}
   (Q_n- Q_{n+1})(1-Q_L) &=& (1-Q_0)(Q_{n-1} - Q_{n}),\\
  \label{Balance2}
   Q_1 (1-Q_L)           &=& (1-Q_0)Q_0    , \\
  \label{Balance3}
   Q_L (1-Q_L)           &=& (1-Q_0)Q_{L-1} , 
\end{eqnarray}
where $n = 1,\ldots,L-1$ in (\ref{Balance1}) and in (\ref{Balance2})
and (\ref{Balance3}) we have taken into account the fact that neither
jumps from a segment containing $0$ particles nor transitions to a
segment with $L$ particles are possible.  Together with (\ref{Prob=1})
and (\ref{Rho}) these relations form $L+3$ equations for $L+1$
variables $Q_n$, $n=0,\ldots,L$, with the concentration $\rho$ being the
only free parameter. This system of equations is easily shown to have
a unique solution
\begin{equation}
  Q_n = \frac{a^n}{1+a+\ldots+a^L},
\end{equation}
where the parameter $a$ can be determined using
\begin{equation}
  \label{Conda}
   L\rho = \frac{\sum_{n=0}^L n a^n}{\sum_{n=0}^L a^n}.
\end{equation}
The concentration $\rho$ is a monotonic function of $a$,
increasing from 0 for $a=0$ to 1 in the limit  $a\to\infty$. The value
$a=1$ corresponds to $\rho = \case{1}{2}$ and, generally, 
\begin{equation}
\label{ro-a}
  \rho(a) = 1-\rho(1/a).
\end{equation}

Having obtained $Q_n$ we can calculate the current as
\begin{eqnarray}
 \label{JBLeft}
J &=& b^{-1}\gamma^4 (1-Q_0)(1-Q_L) \nonumber \\
  &=& b^{-1}\gamma^4 
      \frac{a \left( \sum_{n=0}^{L-1} a^n\right)^2}{
              \left( \sum_{n=0}^{L} a^n\right)^2}.
\end{eqnarray}
Using (\ref{ro-a}) it is easy to see that 
\begin{equation}
 J(\rho) = J(1-\rho).
\end{equation}
Because for $\rho \ll \case{1}{2}$ Eq. (\ref{Conda}) implies $a \approx
\rho/L$, using our formula (\ref{JBLeft}) we conclude that for small
concentrations of particles the current $J$ grows linearly with $\rho$,
\begin{equation}
  J \approx  b^{-1}\gamma^4 L^{-1} \rho.
\end{equation}
For $\rho = \frac{1}{2}$ the mean-field theory (\ref{JBLeft}) predicts
\begin{equation}
  \label{Eq_34} 
  J(0.5) =  \frac{b^{-1}\gamma^4 L^2}{(L+1)^2}\:.
\end{equation}

\subsection{Numerical simulations}

In our simulations we used a lattice with $N=400$ sites consisting of
$\mu=100$ segments, each of length $L = 4$. We used a sawtooth
potential with $\gamma = \exp(-2) \approx 0.135$. The number
of particles in the system varied from $M=1$ to $M=399$.  We carried
out our simulations for $t=10^6$ Monte Carlo time steps per particle
and the results were averaged over $10$ different realizations
of the process, which enabled us to estimate the statistical errors.

\begin{figure}
  \epsfxsize3.2in 
  \epsfbox{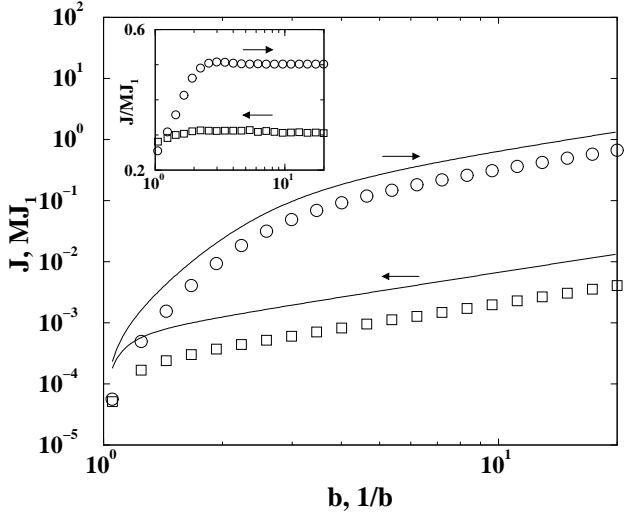}
  \caption{
    \label{Fig3}
Current $J$ (arbitrary units) as a function of the dimensionless bias
parameter for the concentration $\rho = 0.5$. Upper curves: bias to
the right, ordinate indicates $b$; lower curves: bias to the left,
ordinate indicates $1/b$. Full lines: single-particle current $J_1$ of
Ref.\protect \cite{KMWD} multiplied by the number of particles
$M$. Symbols: result of numerical simulations for  $N=400$, $L=4$,
$\gamma = \exp(-2)$, $t=10^6$.  The (semilogarithmic)
inset shows the ratio $J/MJ_1$.
  }
\end{figure}

 We first present simulation results for the current at a fixed
concentration $\rho = 0.5$, or for $M=200$, as a function of the bias
parameter $b$ for bias to the right, and $b^{-1}$ for bias to the
left, respectively. Figure\ \ref{Fig3} shows the current $J$ observed
in simulations (symbols) together with a simple approximation obtained
by multiplying a single-particle current $J_1$ \cite{KMWD} by the
number $M$ of particles in the system (free particle approximation).
One observes that the current
in the case of  a system with a hard-core interactions is reduced as
compared to the
case of non-interacting particles; but the general behavior as a function
of the
bias parameter is practically the same. In particular, the
rectification effects for particle motion in nonsymmetric potentials
are qualitatively the same in both cases. The inset in Fig.\
\ref{Fig3} depicts the ratio $J/M J_1$ as a function of the
bias. Owing to (\ref{J_TASEP}) we expect that for $b\gg 1$ $J/M J_1 =
(N-M)/(N-1) \approx 1-\rho$. For $b=20$ we found  $J/M J_1 \approx
0.5014 \pm 0.0001$, in excellent agreement with the theoretical value
$200/399 \approx 050125$.  For $b \ll 1$ our mean-field approximation (\ref{Eq_34})
predicts $J/M J_1 = L^2/2(L+1)^2 = 0.32$; for $b= 1/20$ our simulations 
yielded a slightly smaller value $0.305 \pm  0.001$.

\begin{figure}
  \epsfxsize3.5in 
  \epsfbox{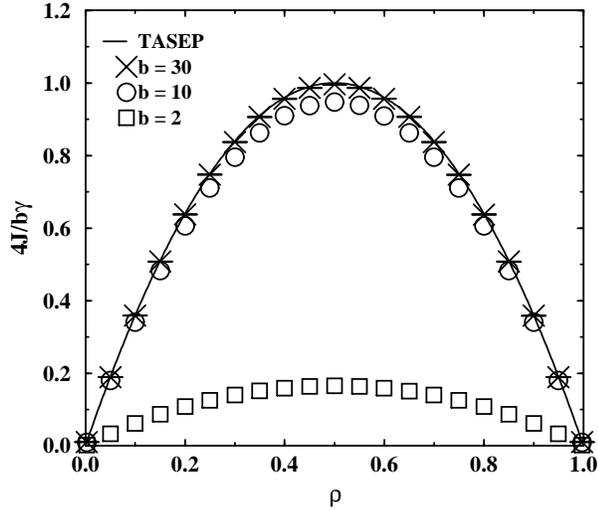}
  \caption{
    \label{FigRight} 
   The dimensionless current, $J/\frac{1}{4}b\gamma$, as a function
    of the dimensionless
    concentration $\rho$ for the bias $b = 30$ (crosses), $10$
    (circles), and $2$ (squares). The solid line was
    computed using Eq.(\ref{J_TASEP_RO}). The parameters are
    $N=400$, $L=4$, $\gamma = \exp(-2)$, $t=10^6$. Results were
    averaged over 10 MC simulations. The error bars are shown only for
    $b=30$; for other values of $b$ they are similarly small.
  }
\end{figure}

We now discuss the dependence of the current on concentration for selected 
values of the bias $b>1$, or $b<1$, respectively, and compare the results
with the theoretical considerations of Sec. IIIB. 
In Fig.\ \ref{FigRight} we present results of our simulations for a
bias to the right ($b=30$, 10 and 2). For a strong bias
($b=30$) the agreement with the theoretical prediction, Eq.\
(\ref{J_TASEP_RO}), is very good.

\begin{figure}
  \epsfxsize3.5in 
  \epsfbox{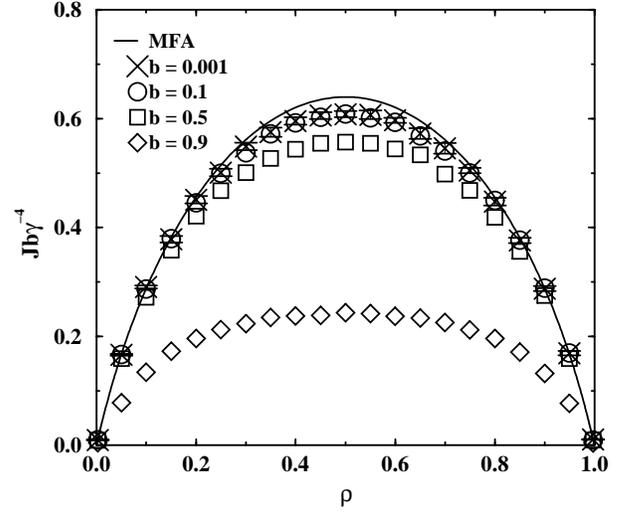}
  \caption{
    \label{FigLeft}  
    The dimensionless current, $J/b\gamma^{-4}$, as a function of the
    dimensionless
    concentration $\rho$ for the bias $b = 10^{-3}$ (crosses), $0.1$
    (circles), $0.5$ (squares) and $0.9$ (diamonds). The solid line
    was computed using Eq.(\ref{JBLeft}). The parameters are
    the same as in Fig.\ref{FigRight}. The error bars, shown
    only for $b=10^{-3}$, are of similar order for other values of
    $b$ and represent the standard deviation.    
  }
\end{figure}

The results obtained for a bias to the left ($b = 0.001$, 0.1, 0.5, and
0.9) are depicted in Fig.\ \ref{FigLeft}. We can see that if the bias
is strong ($b \le 0.1$), the agreement between the mean-field theory
(solid line) and the simulation data (circles and crosses) is very
good for concentrations close to 0 and 1. However, for $\rho \approx
\frac{1}{2}$ we observe that the mean-field theory tends to
overestimate the actual value of $J$ by approximately 5\%, which is
much more than the statistical errors of our data (the relative
standard deviation at $\rho = 0.5$ is about 0.33\%). We repeated our
simulations for larger number of Monte-Carlo time steps ($t =
5\cdot10^6$) and for different values of the bias $b$, but the
difference between simulations and the theory remained practically the
same. We thus conclude that it is not a numerical artifact.  A similar
discrepancy was observed by Tripathy and Barma \cite{TB2}, who
considered a TASEP with random transition rates. However, in their
model the mean-field approach underestimated the magnitude of the
current obtained in simulations for $\rho \approx 0.5$. Moreover, they
found that $J(\rho)$ has quite a broad plateau around $\rho=0.5$. This
phenomenon is not observed in our case because the transition rates in
our model are not random.

\section{Symmetry Properties}

In this section we discuss the symmetry properties of our lattice-gas model with 
nonsymmetric potentials, and of related models. In the simulations, as well as in the 
mean-field approximation, the current exhibits a particle-vacancy symmetry,
\begin{equation}
  \label{sym}
  J(\rho) = J(1-\rho) \:.
\end{equation}
The symmetry properties of the TASEP have been analyzed in \cite{TB2,GS,KP} and the relation 
Eq.(\ref{sym}) has been established in this context. However, the model employed in
those references differs in important aspects from our model. Hence a detailed discussion
is in order.

The particle-vacancy symmetry of the current for the TASEP has been shown in 
Refs.\cite{TB2,GS,KP} for disordered hopping potentials where the transition rates are associated with
the bonds between the sites. If the motion of the particles is reversed (symmetry
operation T
according to Refs.\cite{TB2,KP}, the particles experience the same set of transition rates
as before, only the order of the rates has been changed. If the vacancies are interpreted
as particles (symmetry operation C), they
experience the same transition rates as the particles after the operation T. 
The symmetry under CT is  evident; the nontrivial statement is the symmetry of the current
(up to a sign) under the operations C, or T, separately.

The class of models for the hopping potential that are considered here do not correspond 
to bond disorder. The set of ``right'' transition rates is different from the set of the
``left'' transition rates. If a strong bias $b >> 1$ to the right is applied, leading
approximately to a TASEP, the current is different from the case of strong bias to the
left with $b^{-1} >> 1$. In other words, the symmetry under reversal of motion T does not
exist for the class of models leading to rectification, by their definition. If the 
vacancies are considered as particles, they experience the same set of transition rates
as the original
particles, see also below. We conjecture that symmetry under the operation C also 
exists for our models, if the limiting case of the TASEP is considered. Hence
we expect Eq.(\ref{sym}) to be approximately valid for the models that lead to rectification
effects, in the limit of very strong bias.

The sawtooth potential that is investigated in this paper has a special symmetry which 
will be described now. 
In the limit of concentration of the lattice gas approaching one,
the particle problem is equivalent to the problem of hopping motion of single, independent
vacancies. The hopping transitions of an isolated vacancy are reversed in comparison 
to the transitions of the particle that makes an exchange with the vacancy, e.g.,
\begin{equation}
  \label{gv}
  \Gamma_{l,l+1}^V = \Gamma_{l,l+1} \:.
\end{equation}

\begin{figure}
  \epsfxsize3.2in 
  \epsfbox{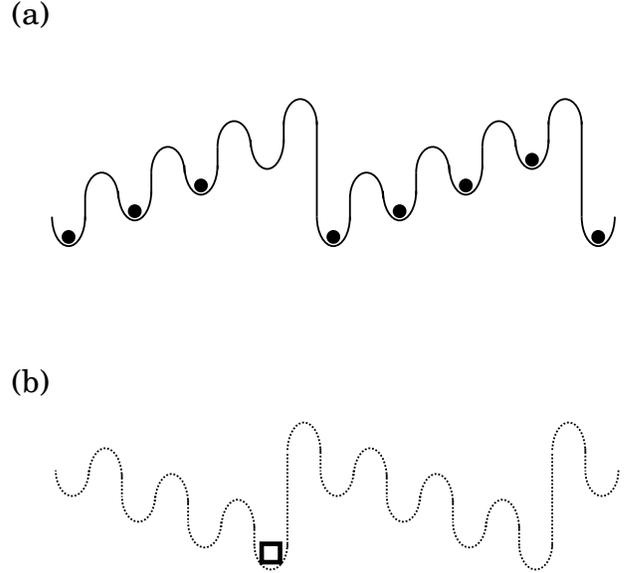}
  \caption{
    \label{Fig6} 
  (a) Repetition of the sawtooth potential of Fig.1 with lattice-gas particles
  indicated. (b) Effective potential for a single vacancy, constructed
  according  
  to Eq.\ref{gv}.
  }
\end{figure}

Using the rates Eq.(\ref{gv}) it is easy to reconstruct the hopping potential for
single vacancies. If this construction is done for the the extended sawtooth potential of 
Fig.1(b), a sawtooth  potential is obtained for the vacancy which is mirror-symmetric with respect
to the original sawtooth potential, see Fig.\ \ref{Fig6}. If a bias is applied to the particles,
expressed by the factor $b$ in the transition rates to the right, the factor $b$ appears
in the transition rates of the vacancy to the left. It is evident from this consideration
that the particle current for $\rho \rightarrow 0$ is identical to the one for
$\rho \rightarrow 1$. 
It is obvious that a particle-vacancy symmetry pertains for the problem
of motion of lattice gases in a sawtooth potential with the above symmetry property;
hence we expect Eq.(\ref{sym}) to be valid for all values of the bias $b$.

We point
out that the sawtooth potential represents a special case; general nonsymmetric potentials
do not provide mirror-symmetric potentials for the vacancies in the limit $\rho \rightarrow 1$.
For instance, if the potential corresponding to an Ehrlich-Schwoebel barrier (see, eg.
\cite{ESB}) is transformed by using Eq.(\ref{gv}) into the corresponding hopping potential
of a single vacancy, a different potential is obtained. As a consequence, the mobility 
of a single particle is different from the mobility of a single vacancy. Hence for this
example $J(\rho) \ne J(1-\rho)$ for $b$ close to $1$. This example is sufficient to show that
the particle-vacancy symmetry (\ref{sym}) cannot be generally valid, for arbitrary $b$.
Another counterexample is provided by the random-trap model, see Ref.(\cite{KPW}).

\section{Concluding remarks}

In this paper we investigated the motion of lattice-gas particles in hopping potentials that
are composed of segments without mirror-reflection symmetry. We considered
in particular the effects of exclusion of multiple occupancy of sites, 
under various bias conditions.
We first studied the case of two particles on a ring of 4 sites with
a sawtooth potential.
The explicit solution of this simple system can be given, and interesting 
conclusions emerge in the limits of large bias to the right, or to the left. We point out
that the ring with 4 sites 
is a model case for the treatment of 2 site-exclusion particles on a finite 
ring; larger systems can be solved in a similar manner, e.g.,  by using symbolic formula 
manipulation programs. 

We then investigated the case of many particles on extended systems which consist of periodic
repetitions of sawtooth potentials.  These systems behave, for strong bias in one direction,
as  uniform systems where the result for the current of lattice gases is known. For
strong bias in the reverse directions, the extended sawtooth potential acts as a periodic
arrangement of weak links. A mean-field expression for the current can be
derived for this case from the cluster dynamics of the particles on the segments, 
which shows similarities 
to the cluster dynamics of the bosonic lattice gases of Ref.\cite{KKRP}. Good agreement 
with the numerical simulations was found for both cases under strong bias; deviations appear
for smaller bias values. The results for the current exhibit a particle-vacancy 
symmetry as a consequence of a special particle-hole symmetry of the 
hopping processes in the sawtooth potential used.

Generally, the current per particle of a site-exclusion lattice gas 
shows the same qualitative behavior  as a function
of the strength and the direction of the bias parameter,  
as the current of independent particles. This observation is important for possible
applications, for instance for transport through channels in membranes or through layered
structures with suitable potential structures. It means that qualitative or even
semiquantitative predictions of the effects of strong bias on the current can already be 
obtained from the single-particle description.

\section*{Acknowledgments} 

We thank G. Sch\"utz for discussions on the TASEP. Z.K. gratefully  acknowledges partial 
support by the Polish KBN Grant Nr 2 P03B 059 12.

\end{document}